\documentclass[12pt,preprint,a4paper]{aastex}
\usepackage{graphics,epsf}
\usepackage{amsmath}                
\usepackage{amsfonts}               
\usepackage{amssymb}                
\usepackage{epsfig}                 
\usepackage{subfigure}
\usepackage{float}
\usepackage{color}

\def \s{~\rm{s}}
\def \km{~\rm{km}}

\def \AU{~\rm{AU}}

\def \yrs{~\rm{yrs}}
\def \yr{~\rm{yr}}
\def \pc{~\rm{pc}}

\begin{document}

\title{The fraction of type Ia supernovae exploding inside planetary nebulae (SNIPs)}

\author{Danny Tsebrenko\altaffilmark{1} and  Noam Soker\altaffilmark{1}}

\altaffiltext{1}{Department of Physics, Technion -- Israel
Institute of Technology, Haifa 32000, Israel;
ddtt@tx.technion.ac.il; soker@physics.technion.ac.il.}
\begin{abstract}
Using three independent directions
we estimate that the fraction of type Ia supernovae (SNe Ia) exploding inside planetary nebulae (PNe),
termed SNIPs,
is at least $\sim 20 \%$.
Our three directions are as follows.
$(i)$ Taking the variable sodium absorption lines in some SN Ia to originate
in a massive circumstellar matter (CSM), as has been claimed recently,
we use the results of Sternberg et al. (2014) to imply that $\gtrsim 20 \%$ of
SN Ia occur inside a PN (or a PN descendant), hence classify them as SNIPs.
$(ii)$ We next use results that show that whenever there are hydrogen lines
in SN Ia the hydrogen mass in the CSM is large $\ga 1 M_\odot$, hence the explosion is a SNIP.
We make the simplest assumption that the probability for explosion is constant in
time for up to about $10^5 \yrs$ after the merger of the core with the white dwarf (WD)
in the frame of the core-degenerate scenario.
This results with at least few $\times10\%$ of SNe Ia that may have a SNIP origin.
$(iii)$ We examine the X-ray morphologies of 13 well-resolved close-by SN remnants (SNRs) Ia
{{{{ and derive a crude upper limit, according to which }}}} $10-30\%$ of all SNRs Ia possess opposite ear-like features,
which we take as evidence of SNIP origin.
Our results, together with several other recent results, lead us to conclude
that the two scenarios most contributing to SNe Ia are the core degenerate and
the double degenerate scenarios.
Together these two account for $>95 \%$ of all SNe Ia.
\end{abstract}

\section{INTRODUCTION}
\label{sec:intro}

It is agreed
that Type Ia supernovae (SN Ia) are
thermonuclear explosions of white dwarfs (WDs) accompanied by a
complete destruction of the WD, or at least one of the two
interacting WDs {{{(e.g. \citealt{Maozetal2014})}}}.
On the other hand, there is no consensus yet on
the evolutionary route of the exploding WD. The different
scenarios discussed in the literature in recent years can be
classified according to different criteria, one of which is
summarized in Table \ref{tab:Table1}. 

We list in Table \ref{tab:Table1} 
{{{{ the relevant observational properties of each scenario that might be compatible with the presence of Ears,
or on the contrary be in conflict with the presence or Ears.}}}}
{{{{  We also }}}} list what we consider as main predictions, strong characteristics, 
and main difficulties of each scenario, {{{{  as these can be relevant to the properties related to Ears.
For example, the presence of massive, approximately $1 M_\odot$ of CSM in the Kepler SNR,
poses a problem to several scenarios, such as the massive CSM inferred in the SN~Ia PTF11kx. }}}}
All scenarios suffer from difficulties, but we find that three
scenarios suffer from severe difficulties,
making them highly unlikely to be the main evolutionary route of SN Ia.
As all scenarios have difficulties,
we list them and discuss them in alphabetical order.
\begin{table*}
\scriptsize
\begin{center}
  \caption{Confronting five SN Ia scenarios with observations}
    \begin{tabular}{| p{1.8cm} | p{2.5cm}| p{2.5cm}| p{2.5cm}| p{2.5cm} | p{2.5cm} |}
\hline  
{Scenario$^{[1]}$}  & {Core Degenerate}    & {Double Degenerate} & {Double Detonation} & {Single Degenerate} & {WD-WD collision} \\

\hline  

 {Presence of 2 opposite Ears
in some SNR~Ia.}
  & Explained by the SNIP mechanism (studied in this paper).
  & Low mass Ears if jets during merger \citep{TsebrenkoSoker2013}.
  & No Ears are expected for He WD companion. 
  & Ears by jets from accreting WD \citep{TsebrenkoSoker2013}.
  & No Ears are expected   \\

\hline  

 {$\approx 1M_\odot$ CSM in Keplers SNR }
  & The massive CSM shell might be a PN.
  & No CSM shell
  & Any CSM is of a much lower mass
  & Can be explained by heavy mass loss from an AGB donor. $^{[5]}$
  &  No CSM shell   \\

\hline  
 {Main 
 
 Scenario
 
 Predictions}
 & 1. Single WD explodes \newline2. Massive CSM in some cases (SNIP)
 & 1. Sufficient WD-WD close binaries \newline2. DTD~$\propto t^{-1}$
  & 1. Asymmetrical explosion \newline2. $M_{\rm WD} <1.2 M_\odot$
  & 1. Companion survives \newline2. $M_{\rm WD} \simeq M_{\rm Ch}$
  &  Asymmetrical \newline explosion   \\
\hline  
 {General
 
  Strong
 
 Characteristics}
 & {\textcolor[rgb]{0.00,0.59,0.00}{1. Explains some SN Ia with H-CSM \newline 2. Symmetric explosion}}
 & {\textcolor[rgb]{0.00,0.59,0.00}{Explains very well the delay time distribution (DTD)}}
 & {\textcolor[rgb]{0.00,0.59,0.00}{Ignition easily \newline achieved}}
 & {\textcolor[rgb]{0.00,0.59,0.00}{1. Accreting massive WDs exist \newline 2. Many explosions with $\sim M_{\rm Ch}$ }}
 & {\textcolor[rgb]{0.00,0.59,0.00}{1. Ignition easily achieved \newline 2. compact object}} \\
\hline  
{General

 Difficulties}
 & {\textcolor[rgb]{0.8,0.0,0.8}{More work on \newline 1. Ignition process\newline 2. DTD\newline 3. Merge during CE \newline 4. Find massive single WDs}}
 & {\textcolor[rgb]{0.8,0.0,0.8}{1. Ignition process \newline2. Inflated gas around merger product$^{[2]}$ \newline 3. Asymmetrical explosion  }}
 & {\textcolor[rgb]{0.8,0.0,0.8}{Ejected He in some sub-scenarios}}
 & {\textcolor[rgb]{0.8,0.0,0.8}{1. Cannot account for DTD \newline2. CSM of PTF 11kx too massive}}
 & {\textcolor[rgb]{0.8,0.0,0.8}{Does not reproduce the DTD at \newline $t<0.4$~Gyr}}
                               \\
\hline  
{Severe

 Difficulties}
 & {\textcolor[rgb]{0.98,0.00,0.00}{}}
 & {\textcolor[rgb]{0.98,0.00,0.00}{}}
 & {\textcolor[rgb]{0.98,0.00,0.00}{1. $M_{\rm WD} < 1.2M_\odot$ \newline2. Highly asymmetrical explosion}}
 & {\textcolor[rgb]{0.98,0.00,0.00}{1. Not enough systems \newline2. Companions not found \newline3. No hydrogen observed}}
 & {\textcolor[rgb]{0.98,0.00,0.00}{1. Can account for $<1\%$ of SN Ia \newline 2.Highly asymmetrical explosion$^{[3]}$  }}
\\
\hline  
{Fraction of

 SN~Ia }
 & $> 20 \% $ (this work)
 & $ < 80 \% $ (this work)
 & $< {\rm few} \times\% $ \newline \citep{Piersantietal2013}$^{[3]}$
 & $0 \% $
 & $<1 \% $ \newline \citep{Sokeretal2014}$^{[4]}$  \\
\hline  
     \end{tabular}
  \label{tab:Table1}\\
\end{center}
\begin{flushleft}
\small Notes:\\
 \small \footnotemark[1]{Scenarios for SN Ia by alphabetical order; see text for details. } \\
 \small \footnotemark[2]{Disk originated material (DOM) around the merger product rules out explosion within several~$\times10 \yr $ of merger \citep{Levanonetal2014}.
 Such a late exploison will solve also the asymmetrical SNR problem.  } \\
 \small \footnotemark[3]{See also \cite{Papishetal2015}.} \\
 \small \footnotemark[4]{See Section 2 in first version of astro-ph (arXiv:1309.0368v1) of \cite{Sokeretal2014}}\\
 \small \footnotemark[5]{ {{{{ In some respects the formation of a shell with Ears in the SD scenario with an AGB star as the donor
 is similar to that in the CD scenario, as both are results of mass loss from an AGB star. }}}} }
\end{flushleft}
\end{table*}

(a){\it The core-degenerate (CD) scenario} (e.g.,
\citealt{Sparks1974, Livio2003, Kashi2011, Soker2011, Ilkov2012, Ilkov2013,
Sokeretal2013}). Within this scenario the WD merges with the hot core
of a massive asymptotic giant branch (AGB) star. In this case the
explosion might occur shortly or a long time after merger.
{{{
By shortly we refer to a time scale within which the SN ejecta will interact with gas originated from the earlier phases of the progenitors,
e.g., the ejected common envelope (CE). By a long time we refer to a time when any mass lost from the progenitor has dispersed in the ISM.
The limit is not sharp, but it is typically $\sim 10^5-10^6 \yr$.
}}}

(b){\it The double degenerate (DD) scenario} (e.g.,
\citealt{Webbink1984, Iben1984}). This scenario is based on the
merger of two WDs. However, this scenario does not specify  the
subsequent evolution, namely, how long after merger does the explosion
of the remnant take place (e.g., \citealt{vanKerkwijk2010}).
Recent papers, for example, discuss violent mergers (e.g.,
\citealt{Loren2010, Pakmoretal2013, Aznar2014}) as possible
channels of the DD scenario. This seems very problematic as it
leads to a highly non-spherical supernova remnant (SNR;
\citealt{Pakmor2012}),
{{{ while \cite{Lopezetal2009} find that SNRs Ia generally possess a spherical structure. }}}
Others consider very long delays from
merger to explosion, e.g., because rapid rotation keeps the
structure overstable {{{ \citep{Piersantietal2003, DiStefanoetal2011, Justham2011, Ilkov2012, TornambPiersanti2013}. }}}
The upper limit on the size of the progenitor of SN~2011fe (e.g.,
\citealt{Nugentetal2011,Bloometal2012,PiroNakar2014}) implies that
the exploding object was very compact, ruling out the presence of
close gas. Since the merger process ejects gas, the DD scenario is
limited to have explosion at a time of $\gg 10 \yr$ after merger \citep{Levanonetal2014}.
As well, the production of manganese \citep{Seitenzahletal2013} and nickel
that requires exploding WDs close to the Chandrasekhar mass
can be explained in the DD scenario if there is enough time for the merging WDs
to coalesce to a massive remnant.

The DD and the CD scenarios can overlap
{{{ in the sense that if the merger occurs after the termination of the CE phase, but while the core is not
yet on the cooling track of a WD, both scenarios comprise the system.
Basically, the scenarios can overlap if the merger occurs during the planetary nebula phase of the system. }}}
However, there are two
significant differences between the two scenarios. In the DD scenario the merger
is of two cool WDs, while in the CD scenario it is a WD with a hot
core. In the DD scenario the delay to explosion is mainly due to
gravitational waves emitted by the two WDs {{{ (e.g., review by \citealt{Maozetal2014}), }}}
while in the CD scenario it is due to angular momentum loss by the merger remnant {{{ \citep{Ilkov2012}. }}}
One significant difference between the two scenarios is that
due to the long time scale of gravitational waves, no massive
circumstellar matter (CSM) is expected around the exploding star
in the DD scenario. On the other hand, in the CD scenario a large
fraction of SN Ia will possess a CSM that will influence the
spectrum of the SN and/or the morphology of the SNR. Since in the
CD scenario this CSM was ionized at some time by the central WD
remnant of the merger, it was a planetary nebula (PN). Such SNe Ia
inside PNe, or remnants of PNe, are termed SNIPs
\citep{TsebrenkoSoker2015}.

(c){\it The double-detonation (DDet) mechanism} (e.g.,
\citealt{Woosley1994, Livne1995}). Here a sub-Chandrasekhar mass
WD accumulates a layer of helium-rich material coming from a donor.
The helium layer is compressed as more material is
accreted and detonates, leading to a second detonation near the
center of the CO WD (\citealt{Shenetal2013} and references
therein). There are two types of He WD companions: A WD of mass
$\sim 0.4-0.45 M_\odot$ residing at $\sim 0.02-0.03 R_\odot$ from
the exploding CO WD \citep{Guillochonetal2010, Raskinetal2012,
Pakmoretal2013}, or a lighter, $\sim 0.2 M_\odot$, WD at an
orbital separation of $\sim 0.08 R_\odot$ \citep{Bildstenetal2007,
ShenBildsten2009}. \cite{Piersantietal2013} found that the DDet
scenario can account for only a small fraction of all SN Ia, but
\cite{Ruiteretal2011} found a much larger fraction that can be
attributed to the sub-Chandrasekhar DDet scenario.
\cite{Papishetal2015} found that the explosion leads to a
non-spherical SNR, and in a case of a close helium WD, the latter will be ignited
and will eject helium.
{{{ As elaborated in \cite{Papishetal2015}, the expected SNR has a dipole asymmetry that no well resolved SNR Ia has.
For example, although SN1006 has a dipole asymmetry, \cite{Papishetal2015} find this dipole to be different from the one expected
due to the effect of the He~WD companion on the ejecta in the DDet scenario. }}}
Another problem with this scenario is that {{{ because the exploding CO~WD almost does not grow before it explodes \citep{ShenBildsten2009}, }}}
it predicts that most exploding WDs will have a mass of $<1.2 M_\odot$.
This is in odd with recent findings that most SN Ia masses are peaked around $1.4
M_\odot$ \citep{Scalzoetal2014}. \cite{Seitenzahletal2013} also
claim that at least $50\%$ of all SN Ia come from near
Chandrasekhar mass ($M_{\rm Ch}$) WDs. We conclude that the DDet
scenario can lead to explosions similar to SN Ia, but not to the
common SN Ia.

(d){\it The single degenerate (SD) scenario} (e.g.,
\citealt{Whelan1973, Nomoto1982, Han2004}). In this scenario a
white dwarf (WD) accretes mass from a non-degenerate stellar
companion and explodes when its mass reaches the Chandrasekhar
mass limit. The search for companions to SNe Ia came with null
results
{{{
(e.g. \citealt{Kerzendorfetal2013, Kerzendorfetal2014})
}}}
and made this scenario very unlikely. Another grave
difficulty is that population synthesis studies find that the SD
scenario can account for only a small fraction of SN Ia (e.g.,
\citealt{Ruiteretal2011, Toonenetal2013, Claeysetal2014}).
{{{
We note that \cite{Hachisuetal2012} who included delay detonation due to WD spin-up, on the other hand, argue that the SD scenario can
produce the required number of SNe Ia, and the observed delay time distribution.
A very interesting set of observations is the detection of CSM from hydrogen lines (e.g., \citealt{Silvermanetal2013, Foxetal2014}) and sodium absorption lines
\citep{Maguireetal2013, Sternbergetal2014} in some SN Ia. These have been used to argue for the SD scenario
(e.g., \citealt{Dilday2012} and \citealt{Silvermanetal2013} from hydrogen, and \citealt{Patatetal2007} and \citealt{Sternbergetal2011} from sodium).
However, it seems that the CSM mass inferred both from hydrogen, as in PTF~11kx \citep{Sokeretal2013}, and from sodium absorption lines
\citep{Soker2014}, is too large to be accounted for by the SD scenario.
Instead, \cite{Sokeretal2013} and \cite{Soker2014} strongly argue that the CD scenario best accounts for these observations.
}}}

(e) \emph{The WD-WD collision scenario} (e.g.,
\citealt{Raskinetal2009, Rosswogetal2009, Thompson2011,
KatzDong2012, Kushniretal2013}). In this scenario two WDs collide
and immediately ignite. The collision is most likely induced by a
tertiary star \citep{KatzDong2012}
{{{
by the Lidov-Kozai mechanism \citep{Lidov1962, Kozai1962}.
}}}
Despite some attractive
features of this scenario, it can account for at most few per cent
of all SNe Ia \citep{Hamersetal2013, Prodanetal2013,
Sokeretal2014}. Another problem is that the explosion in most
cases is highly non-spherical, contradicting the observed morphology of
most young close-by SNRs. A third problem might be the delay time
distribution (DTD) that behaves as $\propto t^{-1}$ (e.g.,
\citealt{GraurMaoz2013}).
{{{
Although the WD-WD collision scenario can lead to the observed $\propto t^{-1}$ DTD, it does so only at late times after star formation,
$>0.4~$Gyr according to \cite{Hamersetal2013}.
The fraction of systems that might be explained by the WD-WD collision is $< 1-2 \%$ of all SN Ia \citep{Hamersetal2013}.
}}}

Based on the discussion above we consider the CD and DD scenarios
as the most likely scenarios to account for most or all SN Ia. In this
paper we try to set a lower limit on the fraction of SN Ia that
come from the CD scenario by considering the influence of the CSM
on the SN spectra (Section \ref{sec:SNCSM}) and on the morphology
of SNRs (Section \ref{sec:SNR}). Our short summary is in Section
\ref{sec:summary}.

\section{SN Ia WITH CIRCUMSTELLAR MATTER}
\label{sec:SNCSM}

\subsection{Sodium absorption lines}
\label{subsec:sodium}

\cite{Sternbergetal2014} find
{{{
that
$19 \pm 12 \%$
of SN Ia in star-forming galaxies exhibit
}}}
time-variable sodium features associated with circumstellar material.
\cite{Maguireetal2013} obtain a similar result,
showing that $\sim 20 \%$ of SNe Ia possess blueshifted narrow Na~I~D absorption features
compared to non-blueshifted Na~I~D features.

Some papers (e.g., \citealt{Patatetal2007,
Patatetal2011, Sternbergetal2011}) attributed the Na~I~D
absorption lines to a wind from a giant star in the SD scenario.
However, it seems that the CSM that supplies the sodium is
too massive to be accounted for in the SD scenario
\citep{Soker2014}. Instead, \cite{Soker2014} suggested that the
sodium comes from dust in a $\sim {\rm few~} \times M_\odot$
shell.
This Na-from-dust absorption (NaDA) model better fits the CD
scenario where the shell is the mass that was ejected in the
common envelope (CE) process (assuming it is not ISM).

For absorption lines to appear, the CSM should reside along the
line of sight. Since in many cases the CSM might not cover the
entire sphere, the $19 \pm 12 \%$ is a lower limit. On the other hand, in
some cases the absorbing gas might be of ISM origin. Overall, the
study of \cite{Sternbergetal2014}, within the frame of the NaDA model,
suggests that $> 20 \%$ of SN Ia have massive CSM that
qualifies them as
{{{
SNIP candidates.
We note that \cite{Shenetal2013}, in the frame of the DDet scenario, and \cite{Raskin2013}, within the frame of the DD scenario,
have also attempted to explain the presence of CSM around some SNe Ia. The CSM mass in these models is too low for the NaDA model \citep{Soker2014},
although can be sufficient for other explanations for the Na absorption lines, e.g. \cite{Patatetal2007} and \cite{Chugai2008}.
}}}

Interestingly, all the SN Ia that show time-variable sodium
features reside in star-forming galaxies \citep{Maguireetal2013}.
This suggests that either the absorbing gas is of ISM origin, as dense clouds are abundant in star forming galaxies,
or that the progenitors of these systems are relatively massive stars, $M \ga 3-4
M_\odot$ that evolve rapidly.
{{{ We assume that the time-variable sodium features are due to CSM as in that case the CSM must reside relatively close to the exploding WD.
In that case the positive correlation of variable absorption with star forming galaxies }}}
is compatible with the CD scenario \citep{Sokeretal2013}.
{{{
We note though that presently the studies of \cite{Sternbergetal2011} and \cite{Maguireetal2013}
cannot be used to distinguish between ISM and CSM features for an individual SN \citep{Sternbergetal2014}.
}}}

\subsection{Massive hydrogen CSM}
\label{subsec:HCSM}

In the last decade more and more SNe Ia have been claimed to interact with a dense CSM
(e.g.,  \citealt{Silvermanetal2013} and \citealt{Foxetal2014}).
We mention here two such SN Ia, SN~2002ic and PTF~11kx.
\cite{Hamuyetal2003} argued that the hydrogen lines in SN~2002ic
can be explained by a few solar masses that have been lost by the
progenitor shortly before the explosion, and attribute it to an AGB
companion in the frame of the SD scenario. \citet{Livio2003}
attributed the massive CSM to the ejection of a CE,
where a WD companion merged with the hot core of a massive AGB
star; a process that was later termed the CD scenario. The massive
CSM was very close, some mass as close as $\sim 100 \AU$, to the SN
origin when explosion occurred.

SN Ia  PTF~11kx \citep{Dilday2012}  had narrow hydrogen lines and
indications of interaction with a massive CSM which started 59
days after the explosion. \cite{Dilday2012} argued that the CSM
was formed by a giant companion to the exploding WD, as is the
case in some channels of the SD scenario. They dismiss the merger
scenario as suggested by \citet{Livio2003} on several grounds,
e.g., it cannot account for several CSM shells.
\cite{Sokeretal2013} argued to the contrary, and explained
PTF~11kx in the frame of the CD scenario. In particular,
\cite{Sokeretal2013} estimated the CSM mass within $\sim 1000 \AU$
to be $\ga 0.2 M_\odot$, much above what the SD scenario can supply.

\cite{Dilday2012} crudely estimate that the  fraction of SNe Ia
that exhibit  prominent  circumstellar interaction  near  maximum
light, e.g., SN~2002ic and PTF~11kx,  is $\sim 0.1 -1 \%$.  As
$\sim 1$--$2$ SN Ia occur per $1000 {\rm M_{\sun}}$ stars formed
\citep{Maozetal2012}, the estimate of  \cite{Dilday2012} stands at
$\sim 0.001  -0.02$ SN Ia with massive CSM per  $1000~{\rm
M_{\sun}}$ stars formed. The Monte Carlo simulations of
\cite{Sokeretal2013} show that the frequency of these systems is
consistent with the CD scenario.

{{{ \cite{Rodneyetal2014} take the SN Ia rate at short delay times of $<500~$Myr
(what they term prompt) to be constant. Based on that, }}}
we take the probability for explosion after the termination
of the CE to be constant with time up to few $\times 10^5 \yrs$.
{{{ In our modelling this rate is higher than that found by \cite{Rodneyetal2014}. Namely, we find in this paper that the
SN Ia rate within about one million years of the CE phase is very high. }}}
For a constant probability per unit time to explode within hundreds of thousands of years after merger,
the probability to explode is linear with the radius of the CSM up to several \pc,
assuming the CSM expands with a constant velocity of tens of $\km \s^{-1}$.
Namely, the probability
{{{
for the system to explode by the time the CSM reached a radius of $\sim 1 \pc$ (as is the case for the Na absorbing gas,)
is $\sim 100$ times that of a system where the CSM only reached a radius of $\sim 0.01 \pc$,
as is the case of the SN Ia-CSM systems.
}}}
Therefore, the $0.1-1 \%$ estimate by \cite{Dilday2012} for cases
where the CSM is close to the exploding site,
translates to  few  $10\%$ of SNIPs (where the CSM is at $\sim 1 \pc$).
This is consistent with the estimate from the Na absorption lines discussed in Section \ref{subsec:sodium}.

\section{SN REMNANTS Ia WITH CIRCUMSTELLAR MATTER}
\label{sec:SNR}

Out of a dozen of resolved SN Ia remnants, at least a few may be
attributed to SN Ia that exploded inside a CSM shell.
For Kepler's SNR, the interaction of the SN Ia with the CSM is established \citep{Patnaude2012}.
Our previous work on this SNR \citep{TsebrenkoSoker2013} suggests
that Kepler's SN exploded inside a CSM shell that originated as a PN.
We term this scenario SN Ia inside PN, or SNIP.
In a recent work, we examined SNR~G1.9+0.3 \citep{TsebrenkoSoker2015}
and suggested that its observed X-ray morphology also hints at a SNIP origin.

Examining the X-ray and radio SNR images in the Chandra SNR
catalogue\footnote{http://hea-www.cfa.harvard.edu/ChandraSNR/}
{{{
\citep{Seward2004},
}}}
we try to identify other SNRs that might have
evidence for SNIP origin.
As an indication for a possible SNIP origin we take here the morphological feature
of two opposite 'ears' in the SNR (nomenclature by \citealt{Cassam-Chenai2004}), a feature both Kepler and G1.9+0.3 have.
On the one hand some `ears' might originate in ISM interaction rather than CSM.
On the other hand some SNIPs might contain no `ears' at all.
As well, we don't consider the possibility of forming the ears by jets
blown by the merging WD and the core (see \citealt{TsebrenkoSoker2013}), although it exists.
For the preliminary estimates of the present study we consider our approach to be adequate.

We focus only on close-by SNRs that have a significant apparent angular size,
and are well-resolved to an extent that allows to identify or
disprove the existence of two opposite ears in the SNR morphology.
All of the SNe in our sample are dynamically young, with estimated ages of less than $10,000 \yr$
{{{{  \citep{ Borkowskietal2006, Caswelletal1982, Hendricketal2003, Hughesetal1995, Parketal2007, Reynoldsetal2008, Vinketal2006}. }}}}
In other words, we only examine SNRs Ia that have clearly
resolved sharp images that can show relatively detailed filaments structure in the SNR shell.
Very old SNRs that have been significantly influenced by ISM interaction are not considered.
Though these criteria limit us to SNRs in our immediate neighborhood (the Galaxy and the Magellanic Clouds; MCs),
we assume that this is a fair representative of other star-forming galaxies.

We have identified a total of 13 SNRs that comply with our criteria and list them in Table \ref{table:SNRs}.
In Kepler's SNR and SNR~G1.9+0.3, and possibly SNRs DEM~L71, G299.2-2.9, 0534-69.9 and 0519-69.0, we identify opposite ears structures.
{{{
We define the opposite ears structure as two protrusions from the overall spherical shape of an SNR,
directly opposite one to another.
The protrusions have width of the order of roughly one tenth of the radius of the SNR so that they can be easily identified by eye.
No other similar protrusions beside these two should be present for a positive identification.
These SNRs are marked as having ears ("Yes") in Table \ref{table:SNRs}.
Some SNRs have such protrusions, but they are less proclaimed or the image is not sharp enough to determine if there are indeed opposite ears.
These SNRs are marked as possibly having ears ("Maybe") in Table \ref{table:SNRs}.
}}}
We already performed hydrodynamical simulations to explain the first two SNRs as SNIPs.
The study of SNR~G299.2-2.9 is a goal of a future work.

From the listed SNRs Ia in the table we derive an initial estimate of $\sim 15-45 \%$ of well-resolved SNRs Ia possessing the opposite-ears morphology.
{{{
In estimating the SNIP fraction other uncertainties should be considered.
First, other scenarios can also create opposite ears, lowering the number of SNIP out of the SNRs Ia having opposite ears.
{{{{ 
Such scenarios include dynamical interaction with the ISM, in which small randomly positioned protrusions can be created by instabilities in the ejecta \citep{Warren2013},
and jets blown during the merger process \citep{TsebrenkoSoker2013}.
Another speculative possibility that we raise here is an asymmetric explosion with jets
along a symmetry axis that can give rise to opposite Ears when the jets interact with the ISM. 
}}}}

We also note that detecting ears in SNRs can be a function of viewing angle.
In \cite{TsebrenkoSoker2013} we estimated that the probability to observe the ears features in Kepler's SNR is roughly $40 \%$.
Therefore, some of the SNRs that we identified as not having ears in our sample might in fact be SNIPs, but with the ears projected in front and behind the SNR itself.
This introduces an additional uncertainty to our crude estimate, lowering the number of SN Ia identified as SNIP candidates.
Another source of uncertainties is the usage of an X-ray catalogue, that might be biased toward SNe Ia exploding in dense ISM and/or within a CSM.
If a SN Ia occurs in a very low density ISM and there is no CSM, then the X-ray emission that results from the ejecta interaction with the surrounding gas will be very weak.
We might miss non-interaction SN Ia. This is likely to be the case with SN Ia in the halo of the Galaxy.
Overall the uncertainties are large, {{{{  and our estimate of the fraction of SNIP candidates in well resolved SNRs Ia being approximately $15-45 \%$ is only a crude upper limit.
}}}}
}}}

The resolved SNRs are all located in the star-forming Milky Way Galaxy and LMC.
\cite{GraurMaoz2013} showed that
star-forming galaxies have a mass-normalized SN Ia rate of $\simeq 0.118$ SNuM $(10^{-12} M_\odot^{-1} \yr^{-1})$,
while passive galaxies (with no recent star formation) have a lower rate of $\simeq 0.082$ SNuM.
Our proposed SNIP scenario is expected to occur in systems that explode through the CD channel (see Section \ref{sec:intro}).
This will predominantly occur in star-forming galaxies.
{{{
\cite{Kelvinetal2014} show that $\simeq 34 \%$ of the stellar mass in galaxies having a stellar mass of $M_\ast > 10^9 M_\odot$ in our local universe ($0.025 < z < 0.06$)
resides within elliptical (passive) galaxies.

Assuming that no SNIPs occur in passive galaxies,
and taking into account the mass-normalized SN Ia rates above \citep{GraurMaoz2013},
and the mass fraction in elliptical galaxies of $34 \%$, 
we find that $\sim 75 \%$ of SN  Ia occur in star forming galaxies.
With our estimate from Table \ref{table:SNRs} of $15-45 \%$ being SNIPs in star forming galaxies,
we arrive to estimate that approximately $10-30 \%$ of all SNe Ia may be SNIPs by the opposite-ears
criterion.
}}}
\begin{table}[H]
        \centering
    \begin{tabular}{|lcc|ccc|}
    \hline
    SNR & X-ray Image & Ears & SNR  & X-ray Image & Ears         \\
    \hline
    Kepler     &  \includegraphics[width=15mm, height=15mm]{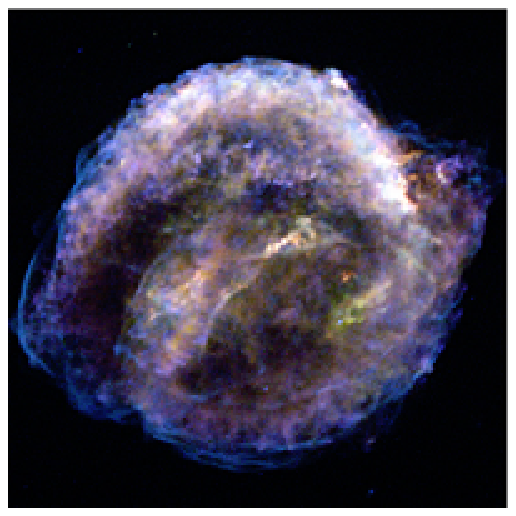}     & Yes      & 1006         &  \includegraphics[width=15mm, height=15mm]{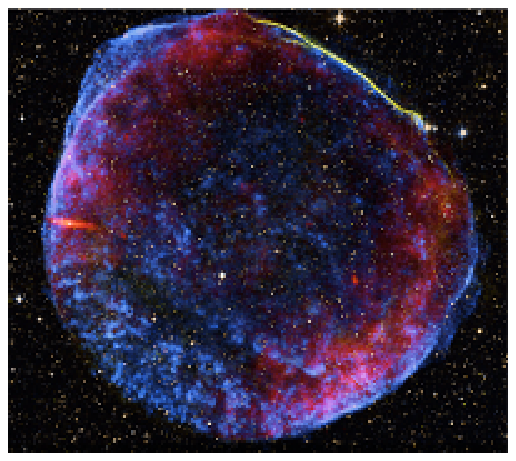}       & No  \\
    G1.9+0.3   &  \includegraphics[width=15mm, height=15mm]{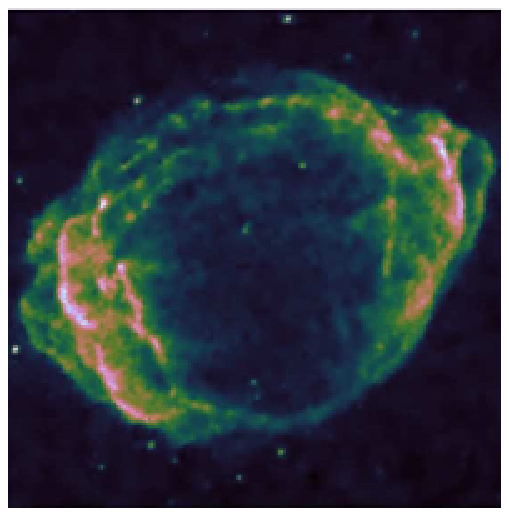}   & Yes      & 3C~397   &  \includegraphics[width=15mm, height=15mm]{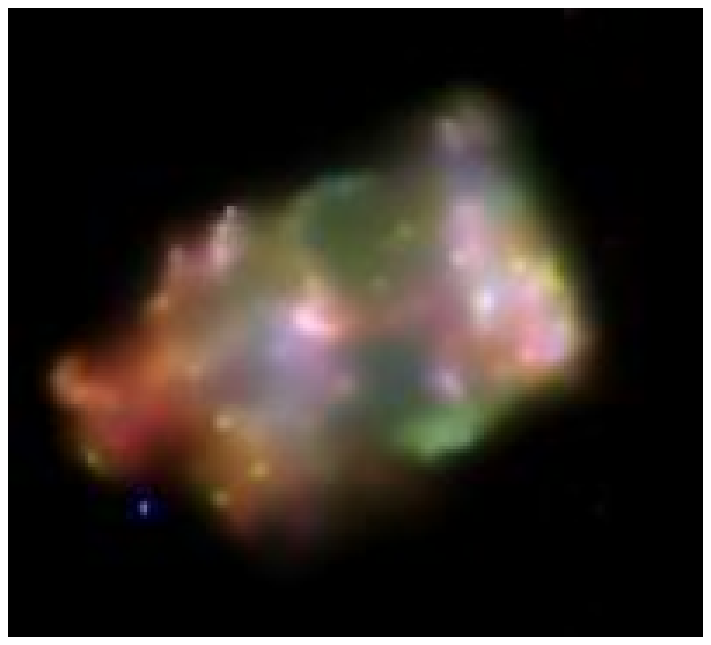} & No   \\
    G299.2-2.9 &  \includegraphics[width=15mm, height=15mm]{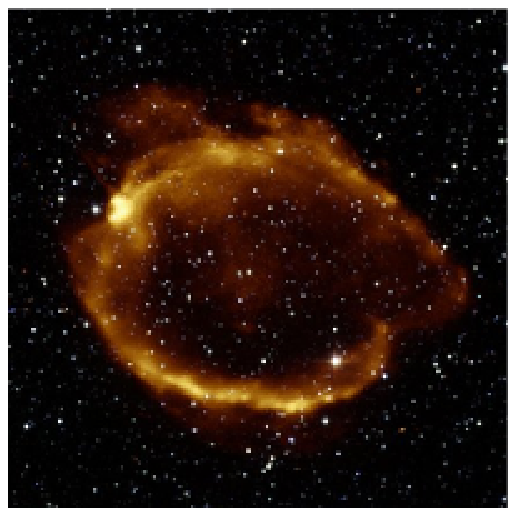}       & Maybe    &  Tycho &  \includegraphics[width=15mm, height=15mm]{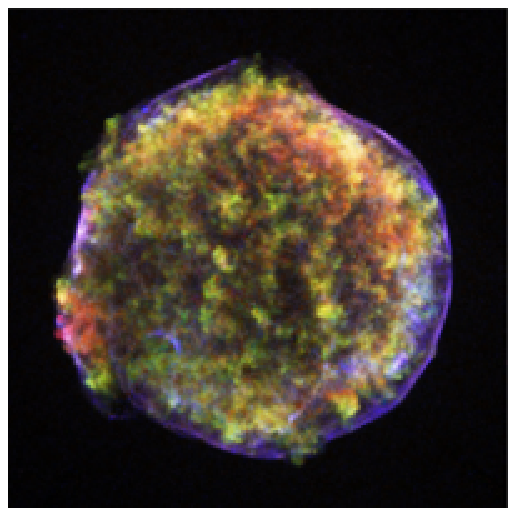}     & No  \\
    RCW86      &  \includegraphics[width=15mm, height=15mm]{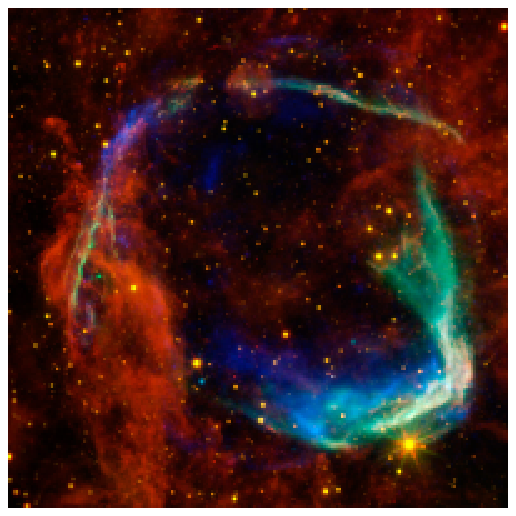}      & No     &   DEM~L71      &  \includegraphics[width=15mm, height=15mm]{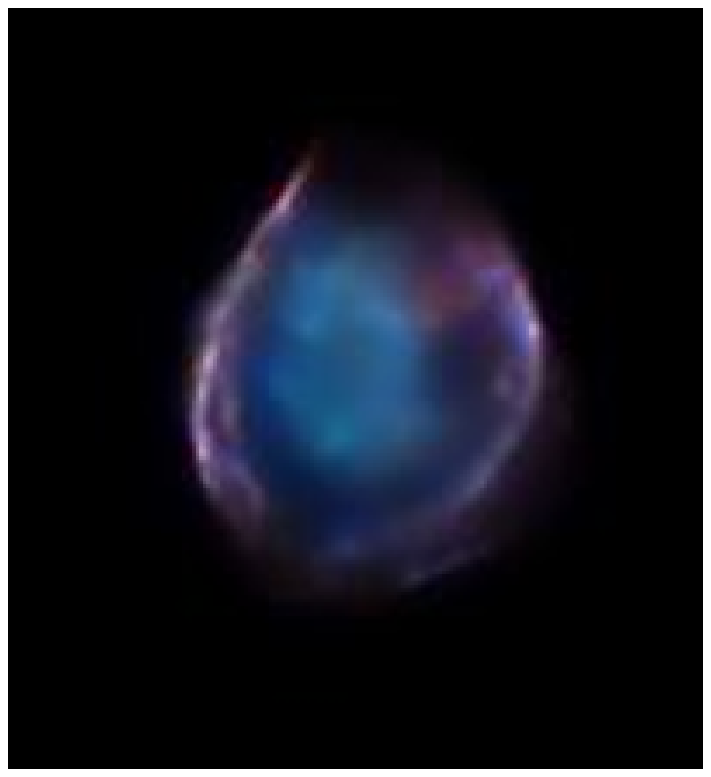}      & Maybe  \\
    N~103B      &  \includegraphics[width=15mm, height=15mm]{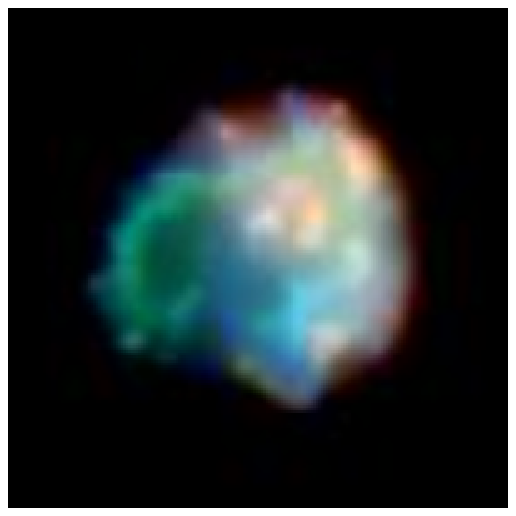}        & No      & 0548-70.4 &  \includegraphics[width=15mm, height=15mm]{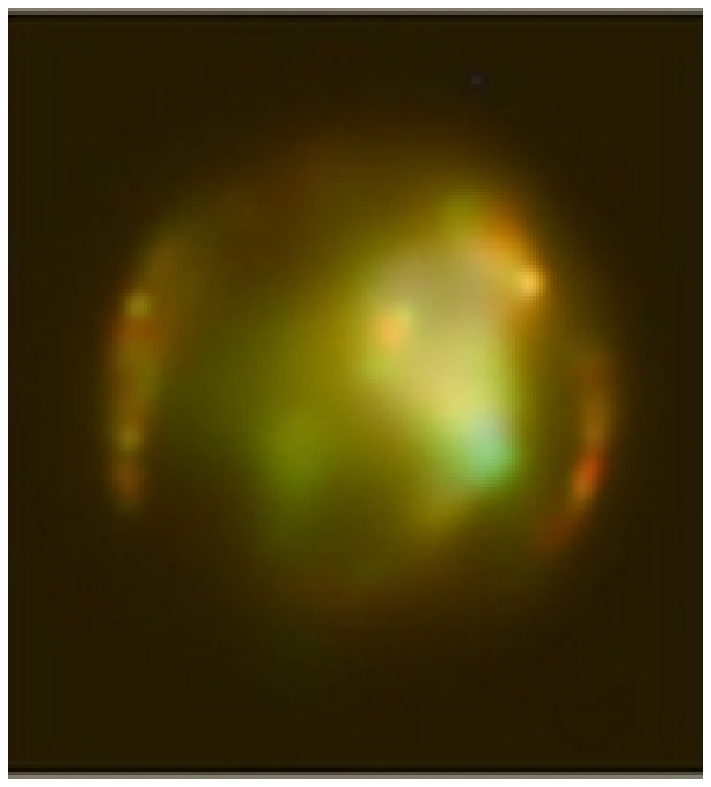}      & No  \\
    0534-69.9  &  \includegraphics[width=15mm, height=15mm]{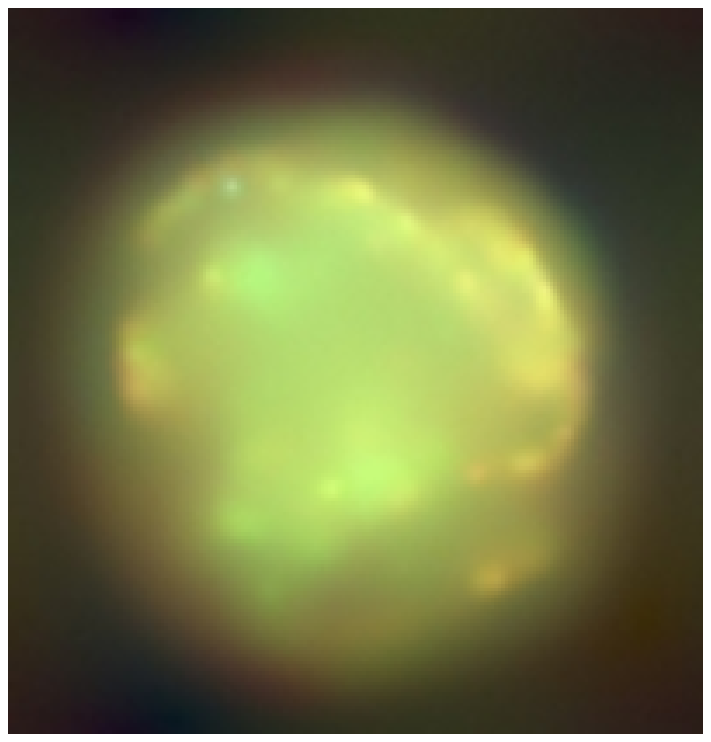}  & Maybe     & 0509-67.5    &  \includegraphics[width=15mm, height=15mm]{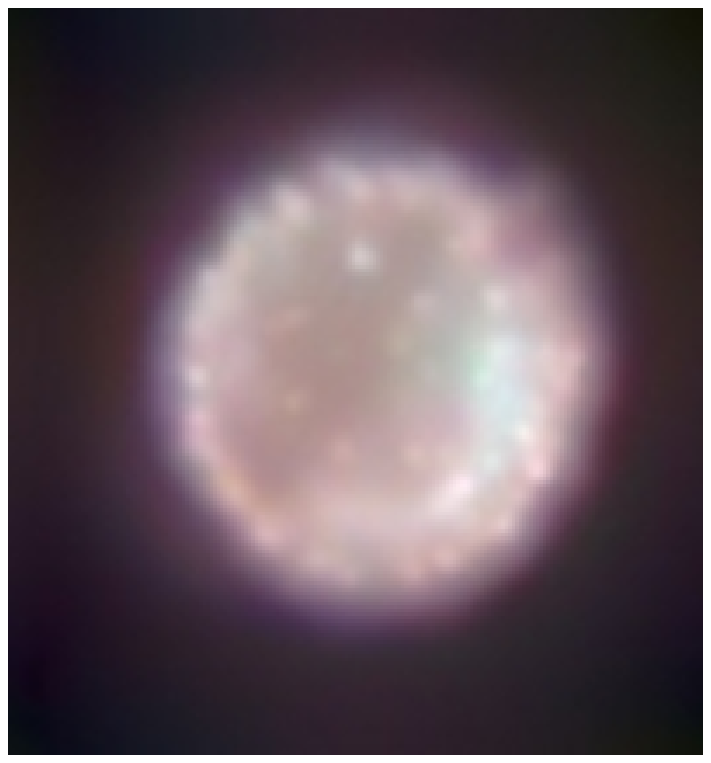}  & No  \\
    0519-69.0  &  \includegraphics[width=15mm, height=15mm]{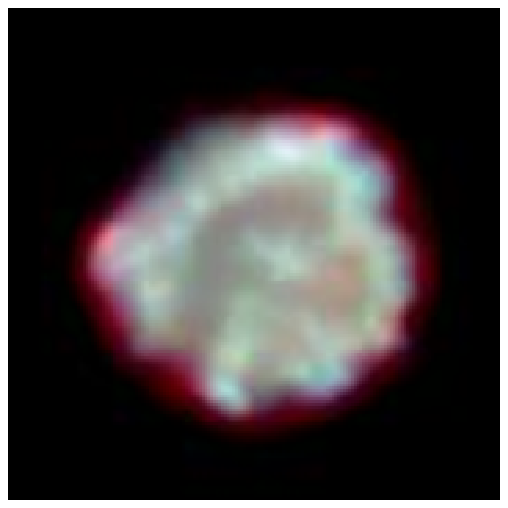}  & Maybe     & -    &  -  & -  \\
    \hline
    \end{tabular}
    \caption{Known well-resolved SNRs {{{ with ages less than $10,000 \yrs$.}}}
             We identify the morphological feature of two opposite ears in between 2 and 6 of the 13 SNRs.
             All images are taken from the Chandra SNR Catalogue (references therein). }
      \label{table:SNRs}
\end{table}

In the CD scenario the two WDs merge at the end of, or very shortly after, the CE phase.
In some cases the two WDs might not merge because the AGB envelope is not massive enough.
The envelope is then ejected before merger.
In such systems, according to the scenario discussed here, the two WDs mass is close or over the Chandrasekhar mass,
and the PN contains "ears", i.e., small lobes.
We are aware of two PNe that have been claimed to host such a WD binary system, and these indeed have "ears".

Nova V458 has a binary system at the center, with a very short orbital period of $\sim 1.63 \rm{hr}$,
consisting of a relatively massive WD ($M_1 \gtrsim 1.0 M_\odot$)
and a secondary post-AGB star ($M_2 \sim 0.6 M_\odot$; \citealt{Rodriguez-Gil2010}).
It is an elongated PN that has two opposite lobes \citep{Corradi2012, Wesson2008}.
The gravitational wave inspiral time for this binary is $\sim 4 \times 10^6 \yr$,
making it too long for a SNIP explosion to take place, but not by far.

PN G135.9+55.9 (TS01) hosts a binary system with an orbital period of $\sim 3.92 \rm{hr}$ and a
total mass very close to the Chandrasekhar limit \citep{Tovmassianetal2010}.
It is a PN that has two opposite ears \citep{Richeretal2003, Stasinskaetal2010}.

These two systems, we argue, support our SNIP scenario.
They belong to systems just outside the boundary of the parameter space
of the CD scenario where merger has been barely prevented,
most likely due to a too low AGB envelope mass.
Many others WD-core systems with a total mass close to the Chandrasekhar mass
merged at the end of the CE phase and formed SNIPs.
\section{SUMMARY}
\label{sec:summary}

In this work we estimated the fraction of SNe Ia exploding inside PNe (SNIPs), using three independent calculations.
$(i)$ An analysis of sodium absorption lines in SNRs Ia, based on \cite{Sternbergetal2014} and \cite{Maguireetal2013} and described in Section \ref{subsec:sodium}.
The results suggest that $ >20\%$ of SN Ia possess massive CSM and thus can be {{{ potentially}}} qualified as SNIPs.
$(ii)$ An estimation of the probability for SN Ia explosion that will have a massive hydrogen CSM,
based on \cite{Dilday2012} and described in Section \ref{subsec:HCSM}.
This yields an estimate of $\sim {\rm few~} \times 10\%$ of all SNe Ia having SNIP origin.
$(iii)$ An examination of the X-ray morphologies of well-resolved close-by SNRs Ia,
described in Section \ref{sec:SNR}.

{{{ Approximately $15-45\%$ of}}} the well-resolved SNRs Ia possess opposite ear-like features,
which we take as evidence of SNIP origin.
Considering the star-forming nature of the Galaxy and the LMC,
and taking into account the mass-normalized SN Ia rates for star-forming and passive galaxies,
this estimate translates to a fraction {{{ of  $\sim 10-30\%$ of }}} all SNe Ia having SNIP origin, {{{{ as a crude upper limit}}}}.

It is not clear yet whether the core-degenerate (CD) scenario can account for all SN Ia and whether
it can reproduce the DTD (for two opposite views see  \citealt{Ilkov2013} and \citealt{Mennekensetal2012}).
However, it seems quite secure that the CD scenario can explain the large fraction of
SN Ia occurring shortly after star formation,
i.e.,  within $t \la 3 \times 10^8 \yr$, i.e.,
shortly after the two massive stars of a binary system end their evolution
on the AGB \citep{Mennekensetal2012, Sokeretal2013}.
Some studies argued for a very small contribution of the CD scenario.
We attribute their results to two inaccurately treated processes.
First is the mass transfer from the more massive star during its RGB and/or
AGB phases to the still main-sequence companion. \cite{Sokeretal2013} pointed out,
based on the code used by \cite{GarciaBerroetal2012},
that the transferred mass should be higher than usually assumed in population synthesis studies.
The second inaccurately treated process is the CE ejection,
which is too "optimistic"
(in the sense that the common envelope is assumed to be easily ejected) in some population synthesis studies, e.g., \cite{MengYang2012}.

We note that \cite{Maguireetal2013} argue that the CD scenario cannot
account for the CSM they find in their study.
{{{
We strongly disagree with that conclusion. Despite the difficulties and open questions listed in Table \ref{tab:Table1}, we suggest that the CD scenario
is the most likely scenario out of the 5 listed there to account for all the properties of the CSM
(e.g., \citealt{Sokeretal2013, TsebrenkoSoker2013, Soker2014, TsebrenkoSoker2015}).
}}}

Our results, as well as other papers cited here, e.g., that the SD scenario contribution to SN Ia is at best very small,
{{{
suggest that the focus in the open questions of SN Ia progenitors should be shifted from the SD scenario to other scenarios.
}}}
Our view, as summarized in the last row of Table \ref{tab:Table1}, is that the two main SN Ia channels are the DD and CD scenarios.

Though we could only propose crude estimates that still demand a deeper and more thorough analysis,
we are confident in the conclusion that a significant fraction (at least $20 \%$, and possibly more)
of SNe Ia are SNIPs, most likely formed through the CD scenario.

{{{ We thank an anonymous referee for many detailed comments that substantially improved both the scientific content of the paper and the presentation of our results. }}}
This research was supported by the Asher Fund for Space Research at the Technion, and the E. and J. Bishop Research Fund at the Technion.

\end{document}